\documentclass[11pt,oneside,a4paper]{article}

\usepackage[utf8]{inputenc}
\usepackage[T1]{fontenc}
\usepackage{graphicx}
\usepackage{listings}
\usepackage{amsfonts}
\usepackage{geometry}
\usepackage{indentfirst}
\usepackage{subfloat}
\usepackage{url}
\usepackage{listings}
\usepackage{color}
\usepackage{float}
\usepackage{leftidx}
\usepackage{amsmath}
\usepackage{makecell}
\usepackage{hyperref}
\usepackage{graphicx}
\usepackage{subfigure}
\usepackage{graphicx}
\usepackage{float}

\newgeometry{tmargin=2cm, bmargin=2cm, lmargin=2cm, rmargin=2cm}
\title{A~concept of a~measuring system for probe kinesthetic parameters identification during echocardiography examination}
\author{Tomasz Winiarski$^{1}$ \and Paweł Balsam$^{2}$ 	\and 
	Maciej Węgierek$^{1}$ \and
	Sonia~Borodzicz-Jażdżyk$^{2}$ \and
	Wojciech~Dudek$^{1}$ \and
	Konrad~Banachowicz$^{3}$ \and
	Janusz~Kochanowski$^{2}$ \and 
	Michał~Marchel$^{2}$
}

\date{	$^{1}$ Warsaw University of Technology, Institute of Control and Computation Engineering, Nowowiejska 15/19, 00-665 Warsaw, Poland, www.robotyka.ia.pw.edu.pl\\
	$^{2}$ 1st Chair and Department of Cardiology, Medical University of Warsaw, University Clinical Center, Banacha 1a, Warsaw 02-097, Poland\\
	$^{3}$ nomagic Sp. z~o.o., ul. Rakowiecka 36, 02-532 Warsaw, Poland;
}

\begin{document}
	
\maketitle

\begin{abstract}
Echocardiography is the most commonly used imaging technique in clinical cardiology. Due to the high demand for this type of examination and the small number of specialists, there is a~need to support the examination process through telemedicine. Moreover, specialist training can be supported by appropriate simulation systems. For (i) creating tailor-made tele-echo robots, (ii) creating echo system simulators, and (iii) conducting echo examination with local or remote expert assistance, knowledge about echo probe kinesthetic parameters during echocardiography examination is advisable. The article describes the concept of a~measuring system for obtaining such data. 
\end{abstract}

\section{Introduction}
\label{sec:introduction}
Echocardiography, the most commonly used imaging technique in clinical cardiology, has improved the prevention, diagnosis, and management of multiple cardiovascular diseases. Nowadays echocardiography is the first-choice diagnostic imaging procedure in most heart diseases, due to its non-invasive, cost-effective nature with a~high value of provided information~\cite{jep151}. 

National Institute of Public Health -- National Institute of Hygiene in Poland estimated that in Poland in 2013 there were only 5 cardiologists per 100 000 patients~\cite{hpz}. Furthermore, in the near future, the problem of Polish healthcare will be aging of cardiologists -- in 2013 the average age of cardiologists in Poland ranged between 40-45 years old~\cite{moh}. Some new technologies improve the diagnosis of cardiovascular diseases and resolve daily clinical problems including poor patients' compliance~\cite{oculus,CardiolJ2018}. Furthermore, one of the best solutions for facilitating access to qualified healthcare is telemedicine. It enables the provision of remote medical services through modern technology. In case of echocardiography, robots can be used to perform remote diagnosis~\cite{Arbeille2014TelesoperatedEU,Boman2009RemotecontrolledRA,Boman2014RobotassistedRE}. On the other hand, the answer to a~small number of specialists are simulation systems that can be used in echocardiography education~\cite{lewiss2014point,arya2017effectiveness,madsen2014assessment}.

Knowledge about echo probe kinesthetic parameters during echocardiography examination are needed for (i) creating tailor-made tele-echo robots, (ii) creating echo system simulators, and (iii) conducting echo examination with local or remote expert assistance. Our article describes the concept of a~measuring system for identification forces/torques (wrenches) exerted by the echocardiography transducer on the patient's body as well as probe trajectory during the examination. Despite the existence of similar solutions~\cite{noh_handle,sandoval_handle,salcudean_handle}, no one has proposed a~comprehensive solution for echocardiography yet.

The paper structure is as follows. In section~\ref{sec:assumptions} the problem's general assumptions are presented. Section~\ref{sec:examination} describes the principles of echocardiography examination. Section~\ref{sec:proposition} presents the solution concept. The article ends up with the evaluation and final words in section~\ref{sec:evaluation}.

\section{General assumptions}
\label{sec:assumptions}
The functional requirement of the postulated solution is the possibility of registering forces/torques (wrenches) exerted by the echocardiography transducer on the patient's body as well as probe trajectory during the examination. Non-functional requirements have their basis in several meta-assumptions. The solution realisation should be rapid and inexpensive. Moreover, it should be simple to implement and safe for the patient and the operator. Considering the above, it was assumed that the postulated solution should be based on existing echocardiography infrastructure and adapted to typical modern probes.

\section{Echocardiography examination}
\label{sec:examination}
\begin{figure}[H]
	\centering
	\begin{minipage}[c]{0.65\textwidth}
		\centering
		\includegraphics[width=\textwidth]{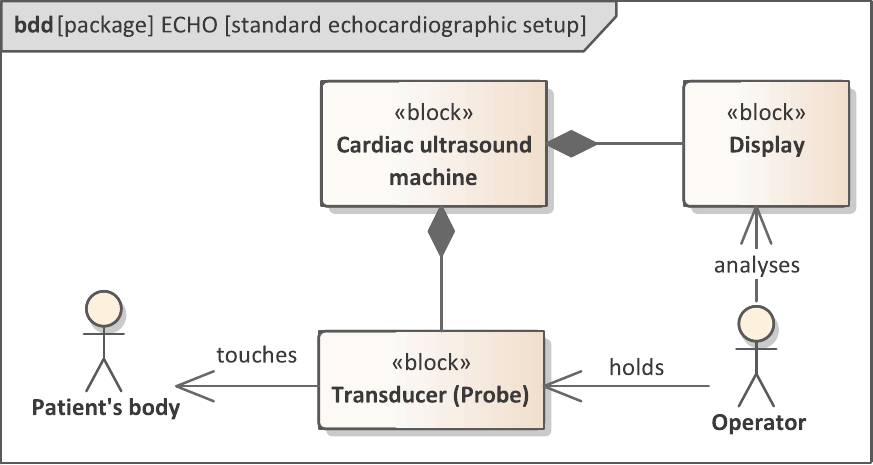}
		\caption{Standard echocardiographic setup}
		\label{img:standard_echocardiographic_setup}
	\end{minipage}
	\begin{minipage}[c]{0.34\textwidth}
		\centering
		\includegraphics[width=\textwidth]{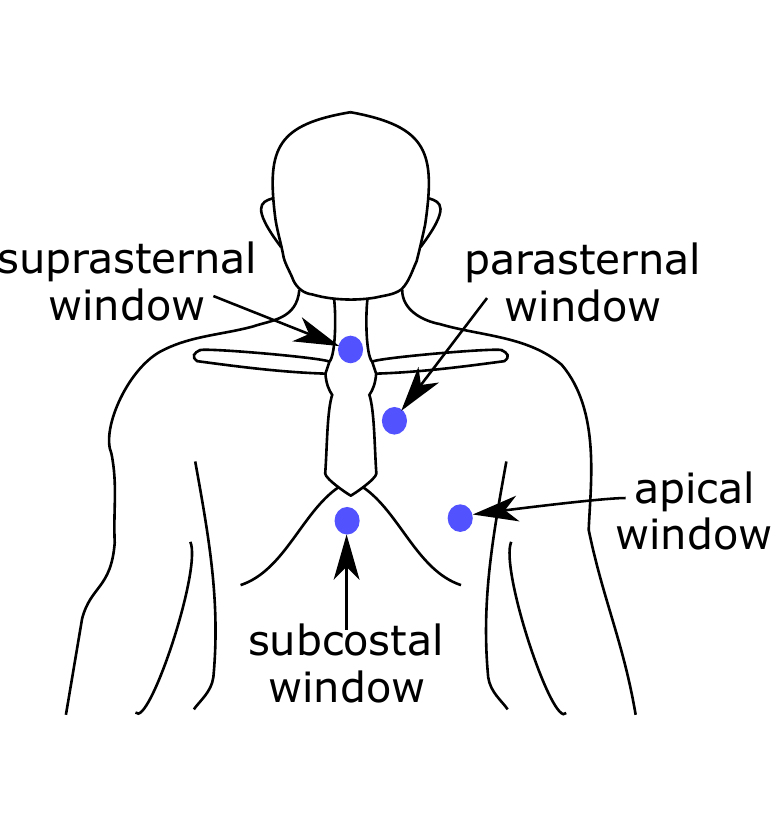}
		\caption{Acoustic windows}
		\label{img:oknaakustyczne}
	\end{minipage}
\end{figure}

To perform the examination, an operator uses an cardiac ultrasound machine of which structure is depicted in a~Fig.~\ref{img:standard_echocardiographic_setup}. The main cardiac ultrasound machine parts are a~transducer (probe) and a~display which shows images. Forces/torques (wrenches) exerted by the echocardiography transducer on the patient's body as well as probe trajectory during the examination are in standard protocol of echocardiographic examination not shown on the display. The operator holds the transducer, which touches the patient's body, and acquires the images in four acoustic windows (Fig.~\ref{img:oknaakustyczne}).

\begin{figure}[H]
	\centering
	\subfigure[]{
		\label{fig:chwyty_g_1g}		
		\includegraphics[height=3cm]{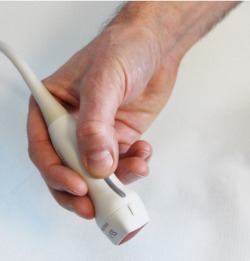}
	}\hspace{1cm}
	\subfigure[]{
		\label{fig:chwyty_g_2g}
		\includegraphics[height=3cm]{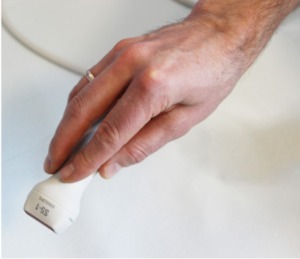}
	}\hspace{1cm}
	\subfigure[]{
		\label{fig:chwyty_g_3g}
		\includegraphics[height=3cm]{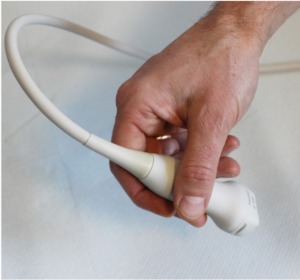}
	}
	\caption{Typical probe grips}
	\label{fig:grips}
\end{figure}

Figure~\ref{fig:grips} presents typical probe grips. The choice of grip depends on the currently examined acoustic window.

\section{Solution design}
\label{sec:proposition}

\begin{figure}[hpbt]
	\centering
	\begin{center}
		\includegraphics[height=6cm]{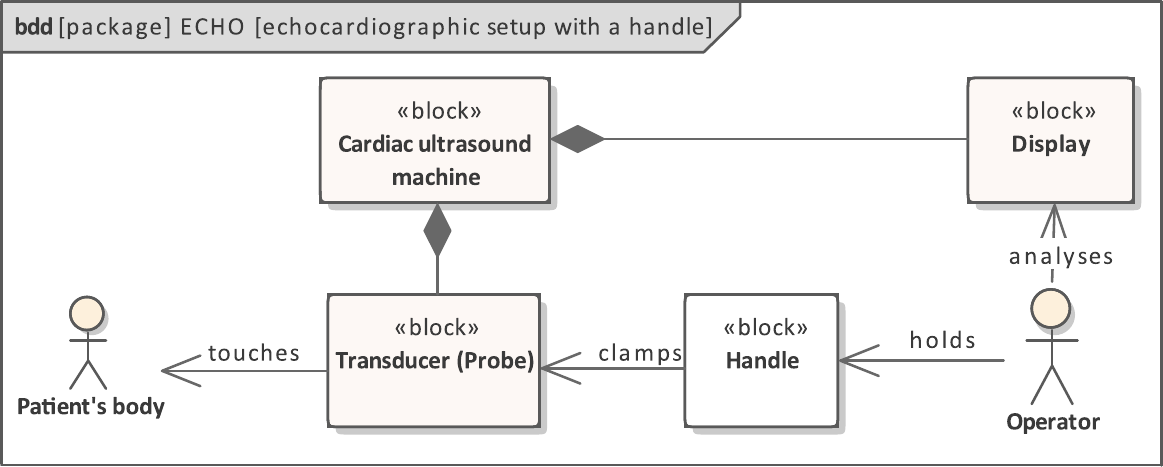}
	\end{center}
	\caption{Echocardiographic setup with a~handle}
	\label{img:echocardiographic_setup_with_a_handle}
\end{figure}

A~concept of a~measuring system for probe kinesthetic parameters identification assumes the use of an additional handle mounted on the probe (Fig.~\ref{img:echocardiographic_setup_with_a_handle}). 

\begin{figure}[h!]
	\centering
	\begin{center}
		\includegraphics[height=8cm]{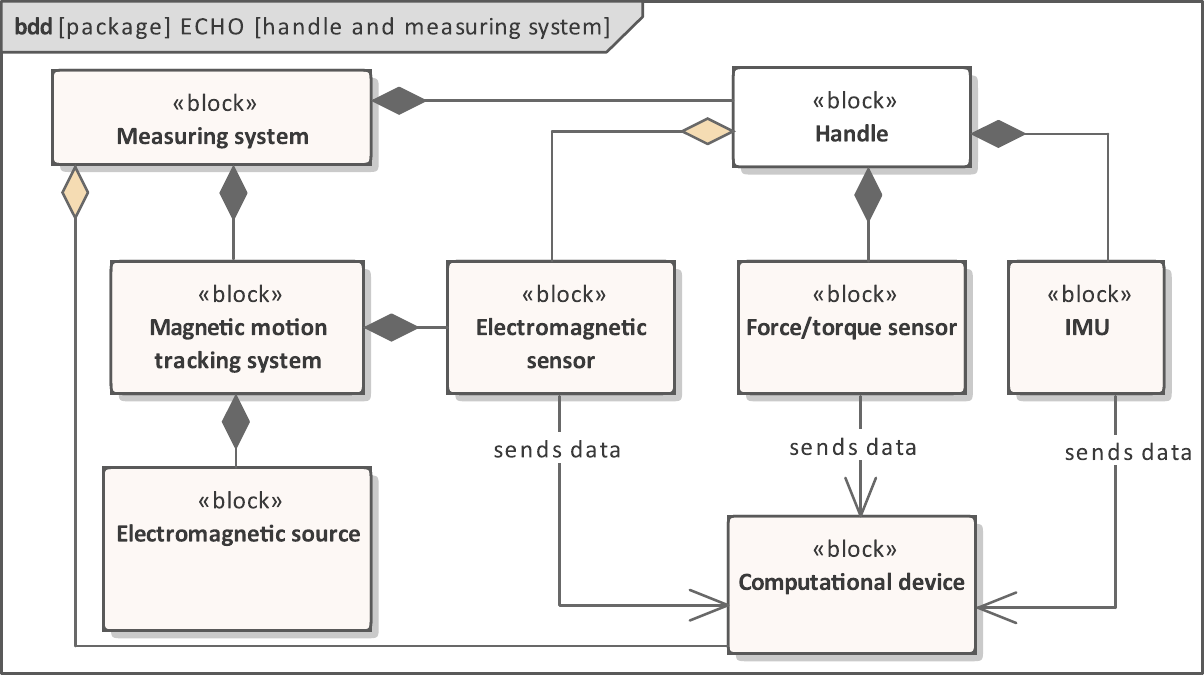}
	\end{center}
	\caption{The measuring system of which the handle is a~part}
	\label{img:handle_and_measuring_system}
\end{figure}

The operator performs the examination similarly as he did when using the basic setup, but in our solution, he holds the transducer through the handle. The handle itself is a~part of a~larger measurement system (Fig.~\ref{img:handle_and_measuring_system}).

Inside the handle, there are several sensors mounted: IMU, force/torque~\cite{ati} sensor, and electromagnetic pose sensor~\cite{pp} that works under the magnetic motion tracking system. The data from these three sensors are sent to a~computational device. The measuring system processes and save data read during the examination.

The force/torque sensor should be used to measure the force/torque exerted by the operator through the probe on the patient body. In order to obtain reliable measurement data, compensation of the gravitational force exerted on the probe with handle should be taken into account~\cite{winiarskirobotica2012,mmar_winiarski_multibehavioral-2016,zielinski2010motion,kornuta-bpan-2020}. The fusion of measurement data from the force/torque sensor and IMU can help to identify the position in which the probe comes into contact with the patient's body~\cite{kroger2007force,kubus2008improving}. A~magnetic motion tracking system can be used for probe trajectory registration. This system is composed of an electromagnetic source that is not a~part of the handle and an electromagnetic sensor that can be mounted in the handle. All three sensors send data to a~computational device that is located outside the handle. The device should, however, be placed fairly close to the handle due to noise reduction of analogue signals and constraints of the wire length for digital signals. 

\begin{figure}[H]
	\centering
	\subfigure[]{
		\label{img:model}		
		\includegraphics[height=5cm]{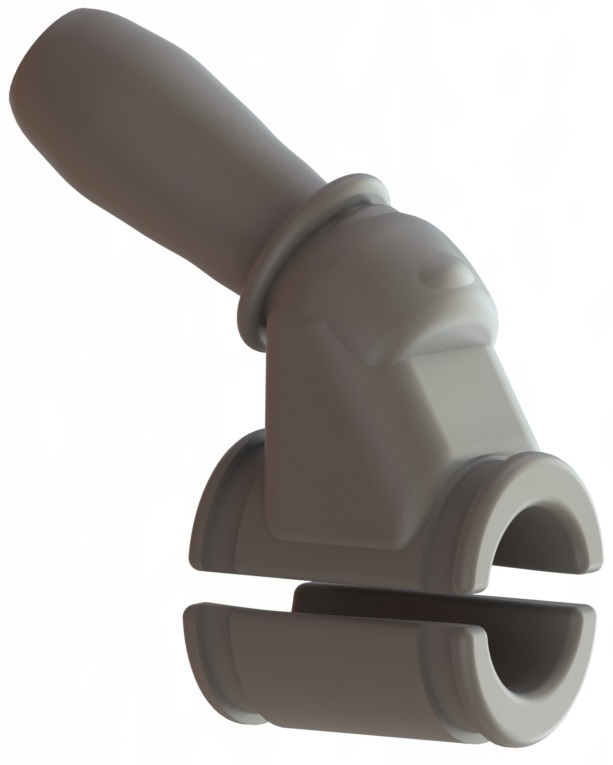}
	}\hspace{3cm}
	\subfigure[]{
		\label{img:wnetrze}		
		\includegraphics[height=5cm]{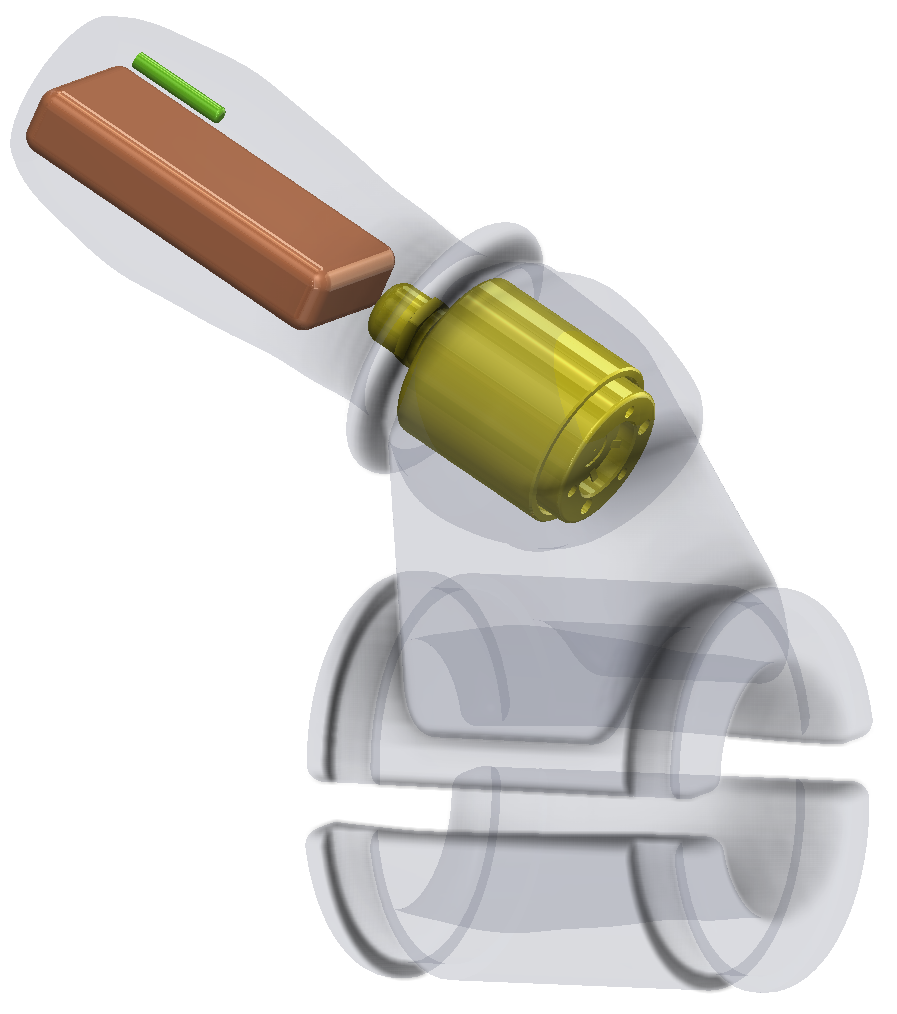}
	}
	\caption{Handle prototype visualisation: (a) -- handle design, (b) -- exemplary handle sensors (yellow -- ATI force/torque sensor $Nano-17 SI-50-0.5$~\cite{ati}, brown -- IMU sensor, green -- POLHEMUS Micro Sensor $1.8^{TM}$ pose sensor~\cite{pp})}
\end{figure}

Our work focuses on the design of the handle and the selection of adequate sensors for the task as well as the concept evaluation. Inside the probe, there is a~space for a~force/torque sensor, IMU, and magnetic field pose sensor. Figures~\ref{img:model} and~\ref{img:wnetrze} present the visualisation of the handle and the sensors. 

\section{Evaluation and final words}
\label{sec:evaluation}
\begin{figure}[H]
	\centering
	\begin{center}
		\includegraphics[width=\textwidth]{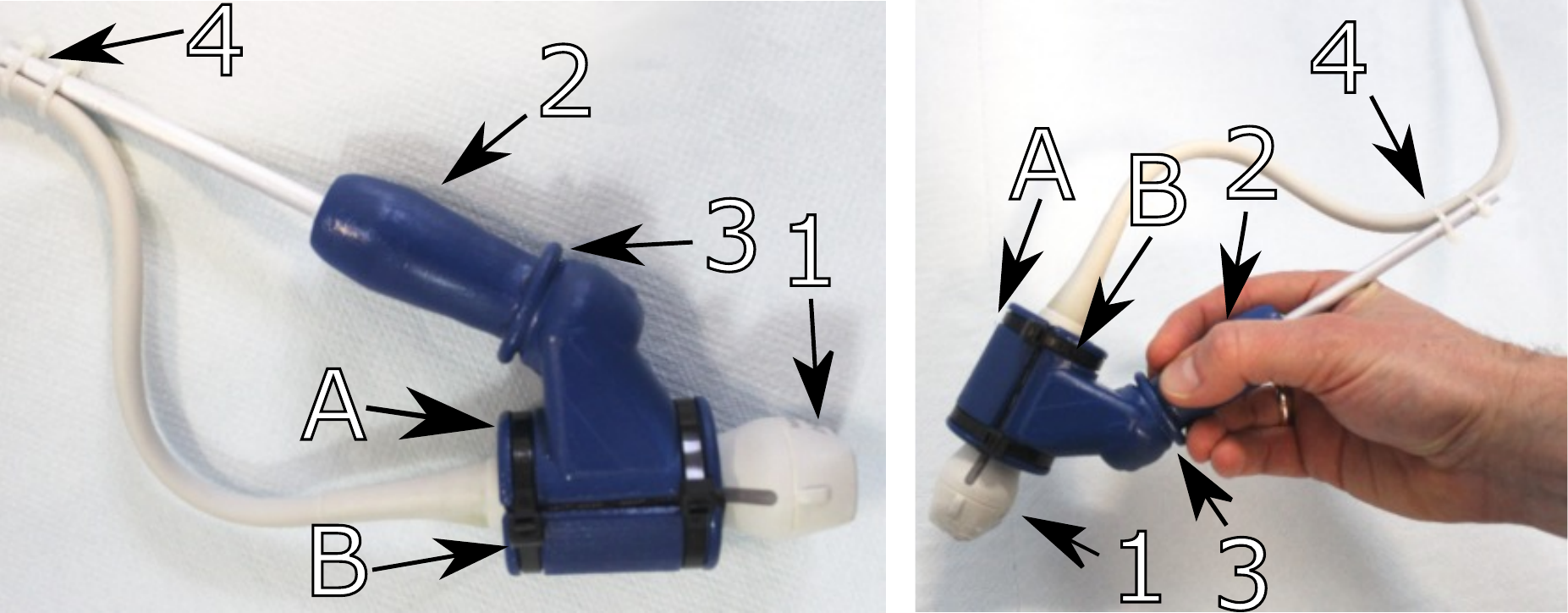}
	\end{center}
	\caption{Physical handle prototype without sensors (two views)}
	\label{img:raczka_numerki}
\end{figure}

We have created a~physical handle prototype without physical sensors that meets the assumptions and limitations. Figure~\ref{img:raczka_numerki} presents 3D printed model of the device. 

The device is formed as a~handle for the echo probe -- a~clamp-like device. It consists of two parts (A,B) which connected together embrace a~probe (1). Due to the fact that the clamp has been properly profiled, the probe is in fixed relation to the clamp. A~layer of soft, rubbery material was placed between the probe and the clamp. Thanks to this, the clamp is better suited to the probe and the use of the clamp does not damage it. An expert holds a~part (2) of the device. This part has a~shape of the echo probe which allows the operator to grip the handle in a~very similar way to gripping a~real probe. A~crossguard ring (3) demarcates an area behind which a~device can not be held by an expert to ensure proper operation of the force/torque sensor. The cable coming out of the probe was attached to a~rigid rod connected to the handle (4). The solution is to prevent a~situation in which this flexible cable, which changes its position during the test, exerts variable forces on the probe. 

A~concept of a~measuring system was evaluated by four experienced and certified cardiac sonographers during a~standard protocol of echo examination in the 1st Chair and Department of Cardiology, Medical University of Warsaw. Forty transthoracic echo exams using the device were performed in both gender patients with variable acoustic window, different chest shape, and body mass index. The device enabled to examine a~patient in a~very similar way as it is usually done using a~standard probe, in all subjects provided standard 2D and Doppler repeatable measurements. 2 out of 4 experts declared the examination using the handle not as comfortable and intuitive as a~standard during echo. However, the average time spent on the exam was comparable to the exam performed using a~routine probe. 

\begin{figure}[h!]
	\centering
	\subfigure[]{
		\label{fig:chwyty_r_1r}		
		\includegraphics[width=0.23\columnwidth]{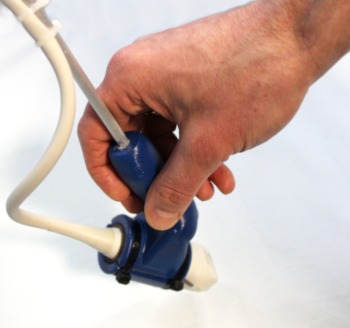}
	}
	\subfigure[]{
		\label{fig:chwyty_r_2r}
		\includegraphics[width=0.23\columnwidth]{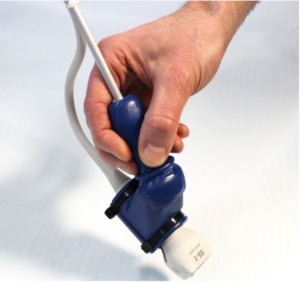}
	}
	\subfigure[]{
		\label{fig:chwyty_r_3r}
		\includegraphics[width=0.23\columnwidth]{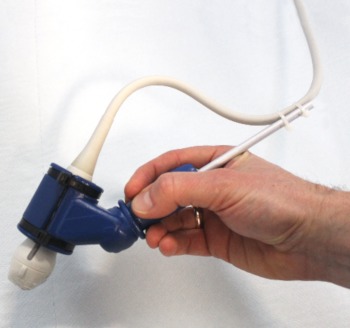}
	}
	\subfigure[]{
		\label{fig:chwyty_r_4r}
		\includegraphics[width=0.23\columnwidth]{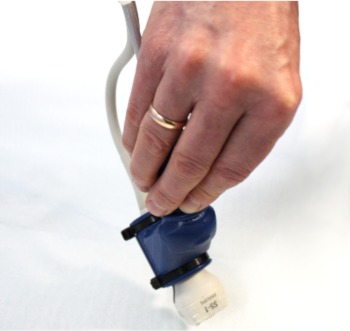}
	}
	\caption{Handle grips}
	\label{fig:chwyty}
\end{figure}

\begin{figure}[h!]
	\centering
	\subfigure[]{
		\label{fig:badanie_1g}		
		\includegraphics[height=2.7cm]{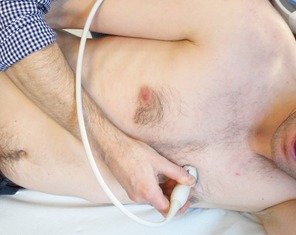}
	}
	\subfigure[]{
		\label{fig:badanie_1r}
		\includegraphics[height=2.7cm]{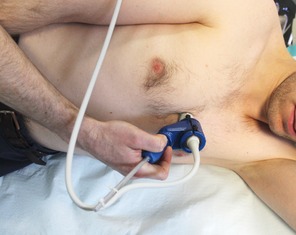}
	}
	\subfigure[]{
		\label{fig:badanie_2g}		
		\includegraphics[height=2.7cm]{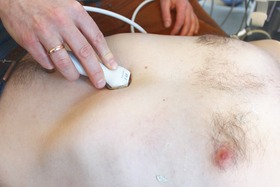}
	}
	\subfigure[]{
		\label{fig:badanie_2r}
		\includegraphics[height=2.7cm]{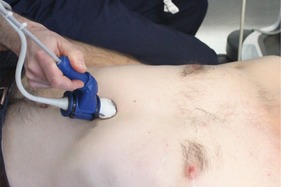}
	}
	
	\subfigure[]{
		\label{fig:badanie_3g}		
		\includegraphics[height=3cm]{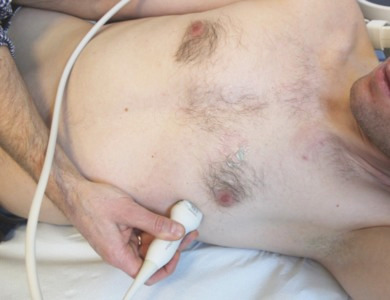}
	}\hspace{0.3cm}
	\subfigure[]{
		\label{fig:badanie_3r}
		\includegraphics[height=3cm]{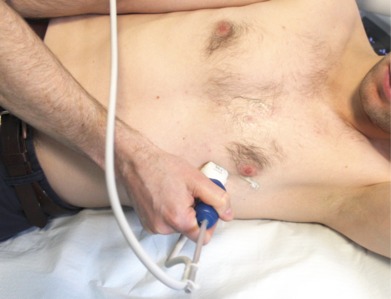}
	}\hspace{0.3cm}
	\subfigure[]{
		\label{fig:badanie_4g}		
		\includegraphics[height=3cm]{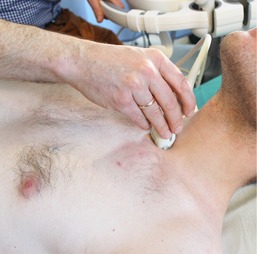}
	}\hspace{0.3cm}
	\subfigure[]{
		\label{fig:badanie_4r}
		\includegraphics[height=3cm]{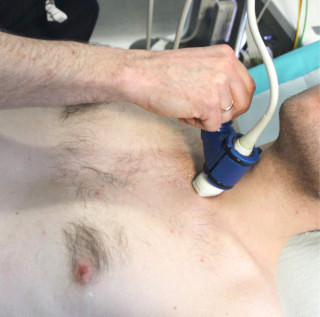}
	}
	\caption{Comparison of echo examination runs with and without the handle -- a,b -- parasternal window; c,d -- subcostal window; e,f -- apical window; g,h -- suprasternal window}
	\label{fig:badanie}
\end{figure}

Figure~\ref{fig:chwyty} shows how the operator can grip the handle. Grips are very similar to those used when holding the transducer directly. Figure~\ref{fig:badanie} compares the performance of the test for four acoustic windows without using the handle and together with the handle.

In our opinion, the above analysis and particular concept evaluation led to the conclusion that creation of a~postulated measuring system is feasible. When equipped with indicated sensors it would allow registering forces/torques (wrenches) exerted by the echocardiography transducer on the patient's body as well as probe trajectory during the examination. Thanks to the fact that the measuring system is based on existing echocardiography infrastructure and adapted to typical modern probes it would be simple to implement and safe for the patient and the operator.

\section*{Acknowledgements}
The authors would like to thank Bartłomiej Kozakiewicz for the help in designing the 3D model of the handle, DPS Software \footnote{https://www.dps-software.pl} company for 
providing SolidWorks software, and 3DMaxBaum \footnote{https://3dmaxbaum.pl} company for making 3D prints of the handle.

	\bibliographystyle{abbrv}
	\bibliography{arxiv-echo-20}
	
\end{document}